\documentclass[aps,twocolumn,prl,tightenlines,floatfix,showpacs,superscriptaddress]{revtex4-1}
\usepackage{graphicx}
\usepackage{amsmath}
\usepackage{amssymb}
\usepackage{mathrsfs}
\usepackage{array}
\usepackage{pdfpages}

\allowdisplaybreaks[4]

\newcommand{\bs}[1]{\boldsymbol{#1}}

\newcommand{\pth}[1]{\left( #1 \right)}
\newcommand{\brc}[1]{\left\{ #1 \right\}}

\newcommand{\abs}[1]{\left| #1 \right|}
\newcommand{\avg}[1]{\left\langle #1 \right\rangle}

\newcommand{\pr}[0]{\partial}

\newcommand{\ria}[0]{\rightarrow}

\newcommand{\meq}[1]{\begin{equation} #1 \end{equation}}

\begin{document}

\title{Phase Imprinting in Equilibrating Fermi Gases: The Transience of
Vortex Rings
and Other Defects}

\author{Peter Scherpelz}
\author{Karmela Padavi\'c}
\author{Adam Ran\c con}
\affiliation{James Franck Institute and Department of Physics,
University of Chicago, Chicago, Illinois 60637, USA}
\author{Andreas Glatz}
\affiliation{Materials Science Division, Argonne National Laboratory, 9700 South
Cass Avenue, Argonne, Illinois 60439, USA}
\affiliation{Department of Physics, Northern Illinois University, DeKalb,
Illinois
60115, USA}
\author{Igor S. Aranson}
\affiliation{Materials Science Division, Argonne National Laboratory, 9700 South
Cass Avenue, Argonne, Illinois 60439, USA}
\author{K. Levin}
\affiliation{James Franck Institute and Department of Physics,
University of Chicago, Chicago, Illinois 60637, USA}

\date{\today}

\begin{abstract}
We present numerical simulations of phase imprinting experiments in ultracold
trapped Fermi gases which are in good agreement with 
recent, independent experimental results.
Our focus is on the sequence and evolution of defects using the
fermionic time-dependent Ginzburg-Landau equation, which contains dissipation necessary
for equilibration.
In
contrast to other simulations we introduce small, experimentally unavoidable symmetry breaking, particularly
that associated with
thermal fluctuations 
and with the phase imprinting tilt angle, and illustrate their
dramatic effects. The former causes
vortex rings in confined geometries
to move to the trap surface and rapidly decay into more stable vortex lines,
as appears consistent with recent experimental claims.
The latter 
aligns the precessing and relatively long-lived vortex filaments, rendering them difficult
to distinguish from solitons.  
\end{abstract}

\maketitle

\textit{Introduction.}
The evolution of defects (solitons or domain walls, vortices, 
cosmic strings, monopoles,
and other topological structures) is central to many sub-disciplines in
physics, including cosmology,
as well as condensed matter and fluid dynamics.
An issue of broad interest is the way in which 
the nature of the defects changes as equilibration ensues, reflecting the
specifics of
the initial perturbation as well as the boundary conditions. 
In trapped superfluid atomic gases one has access to real time dynamics. Here, rapid cooling from the normal phase 
\cite{glatz_2011,weiler_2008},
sudden density
cuts \cite{andrews_1997} and phase imprinting all have led to a 
sequence of defects which may, in some cases, be
relevant to the
Kibble-Zurek mechanism in cosmology as well as to superfluid turbulence.

Phase imprinting is a quite violent alteration of
the superfluid in which part of the superfluid experiences a phase shift 
close to $\pi$.
This
perturbation has attracted recent attention in trapped Fermi gas superfluids
because
of
reports of rather long-lived solitons having anomalously
large effective masses and oscillation periods \cite{yefsah_2013}.
These dark solitons are planar density depletions which maintain their shape 
and are associated with phase changes, often close to $\pi$.
Subsequently,
there has been a more
in-depth
analysis of these experiments \cite{ku_2014} 
which associates
 these
original
solitonic observations with long-lived line vortices.
These so-called ``solitonic" vortices are to be
viewed as vortex filaments in an anisotropic trap.

In this paper we present numerical simulations which relate to these phase
imprinting experiments. 
Our work supports the notion of
a natural hierarchy towards less extended and less complex defects
(soliton to vortex ring to single vortex)
during the processes of equilibration.
One might similarly expect that
the simplest defects are often the most difficult to relax;
indeed, we find that
the residual filamentary vortices 
continue for some time to precess around the atomic cloud.
Additionally we find that
a very small tilt of the phase imprinting angle
is sufficient to consistently align this line vortex in the trapped gas.
The picture that emerges is one that appears generally consistent with
recent experimental reports \cite{ku_2014}. 
The above observations underline a
key contribution of this paper: 
to show how simulations of defect evolution
require one to subject the numerics to checks for robustness against 
unavoidable broken spatial symmetries in the experiment.

Our work should be 
contrasted with previous numerical
simulations  
of the
Gross-Pitaevskii (GP) and Bogoliubov-de Gennes (BdG) equations
which addressed the previously claimed solitonic stability and
attributed it instead to vortex rings (that is, vortex lines which are closed
into a loop)
\cite{bulgac_2014,reichl_2013}.
These past
simulations (omitting noise or other broken symmetries, and in many cases
omitting dissipation) do not appear to exhibit all the transitions
necessary for the full equilibration of the cloud.
Thus the
literature
\cite{bulgac_2014,reichl_2013}
has focused on symmetric vortex
rings which maintain inversion
symmetry about the $z$-axis [parity transformation $(x,y)$ to $(-x,-y)$]
and often apply
initial conditions specifically designed to produce
a stable vortex ring.
In recent work which introduced a small trap anisotropy \cite{reichl_2013}, there were
reports of a vortex ring breaking into two vortex lines. However, in this work
reflection symmetries were still maintained, so that the two lines of opposite
vorticity 
perpetually move along the
longitudinal axis, and after nearing the edge recombine into a ring
moving in the opposite direction. We show that these complex processes,
which do not bring the system closer to equilibration, are not found here and
are overwhelmed by other symmetry breaking features.

\begin{figure}
\includegraphics[scale=0.59]{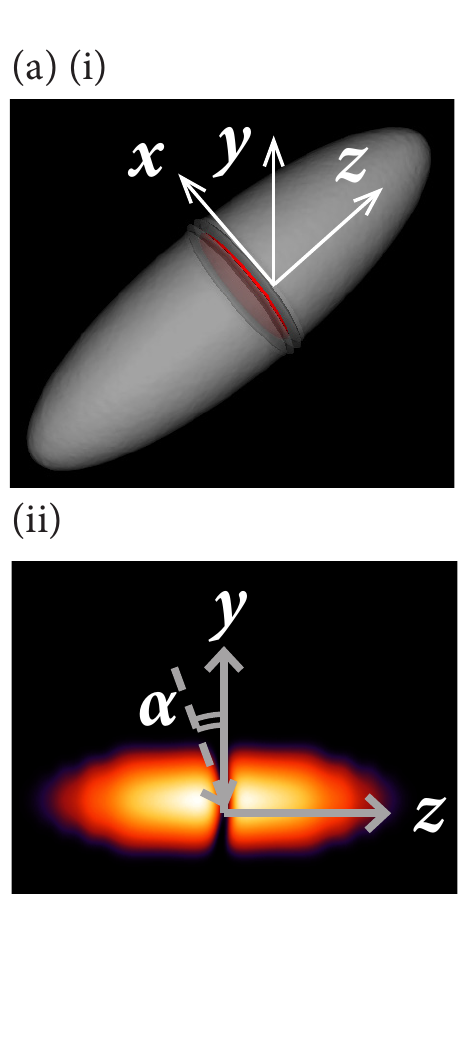}
\includegraphics[scale = 0.59]{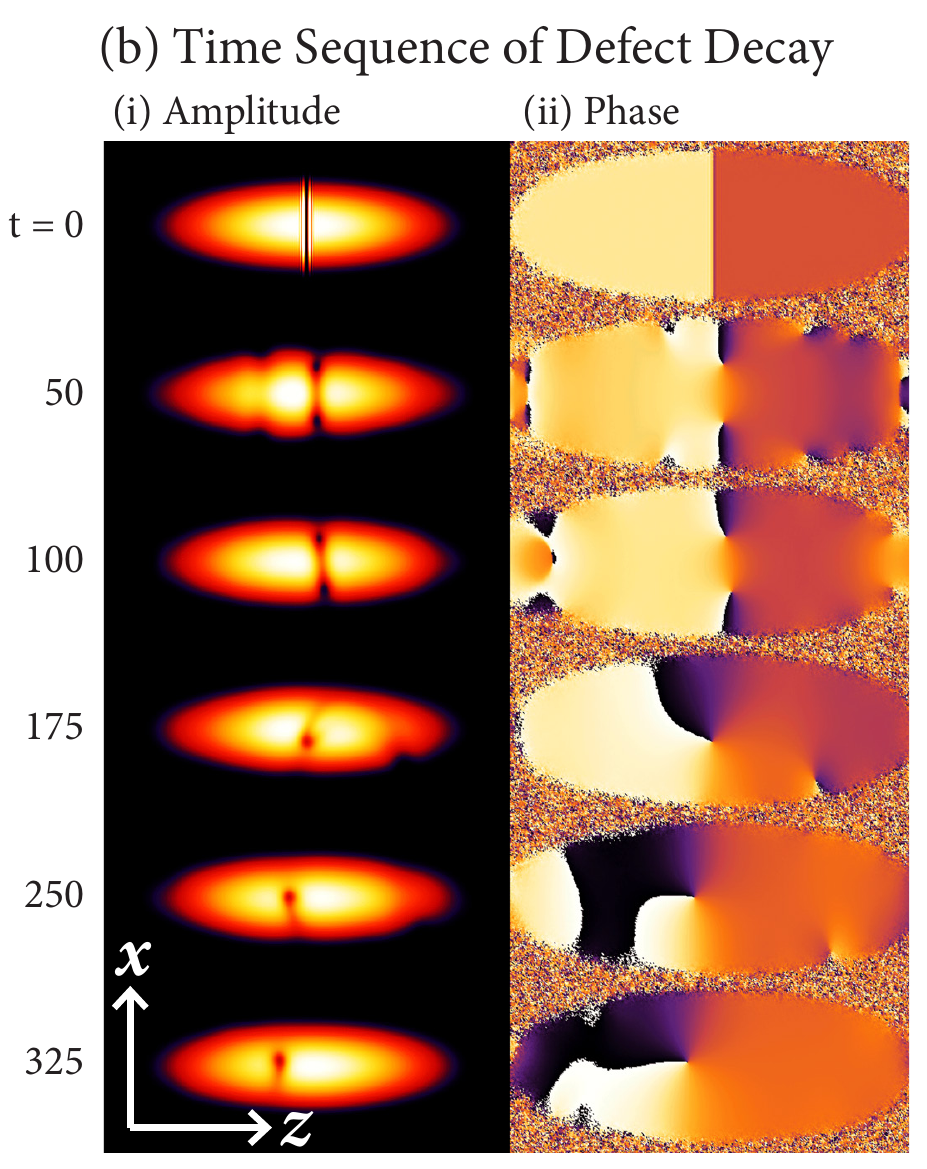}
\caption{\label{fig:r_decay}
    (Color online)
    (a) Depictions of axes used. (i) 
    3D rendering of initial depletion plane (red) with axes used in the
    paper: $y$ is vertical, and $z$ is the long axis of the trap. (ii) Side view
    of axes at $t = 20$ after phase imprinting for the run in (b), 
    along with the tilt of the imaging beam, $\alpha$, 
    considered later in this work. (b)
    TDGL simulation results for one time sequence,
    focusing on the vortex ring decay,
    of a phase imprinting applied at $t=0$ with $\lambda = 3.3$ and parameters
    as in the text.
    (i) shows the cloud density integrated along $y$
    and (ii) the phase in the $y=0$ plane [from 0 (black) to $2\pi$ (white)].
    Here phase imprinting has formed a ring ($t=50$), which is just slightly off
    center. The ring moves toward the cloud edge ($t=100$), then
    impacts the edge and decays to a vortex line ($t=175$). 
}
\end{figure}

\textit{Our approach.}
In contrast to the bosonic GP dynamics, we use
the complex time-dependent Ginzburg-Landau (TDGL) equation  
\cite{aranson_2002} for fermionic superfluids,
to address
three-dimensional (3D)
anisotropic trapped Fermi gases
subjected to a near-$\pi$ phase imprinting quench
\cite{yefsah_2013}.
We consider
a harmonic confining potential
$V(\bs x) = (\omega_\perp^2(x^2+y^2) + \omega_z z^2)/2$, where $\omega_\perp$ and
$\omega_z$ define how tightly
confining the trap is in comparison to the chemical potential, as well as the
trap ratio $\lambda = \omega_\perp/\omega_z$.
Our central TDGL equation is $e^{-i\theta}\pr_t\psi(\bs x,t)=$ 
\meq{ \brc{[1-V(\bs
x)]+\frac{1}{2}\nabla^2-\abs{\psi(\bs x,t)}^2}\psi(\bs
x,t) + \chi(\bs x,t)\label{eq:tdgl},}  
where $\chi$ is uniformly distributed thermal
noise. 
Equation \eqref{eq:tdgl} contains two parameters, in addition to the trap
frequencies. First,
$\theta$ controls the
amount of dissipation, effectively moving from BCS ($\theta = 0$, dissipative)
to BEC ($\theta = \pi/2$, reducing to the time-dependent GP equation, or
equivalently the nonlinear Schr\"{o}dinger equation). 
Second, through
$\avg{\chi(\bs x,t)\chi^*(\bs x',t')} = 2\gamma T_\chi\delta(\bs x - \bs
x')\delta(t - t')$, a fluctuation temperature can be set
\cite{aranson_1999,damski_2010,glatz_2011,schmid_1969}. 
In this work the dimensionless 
fluctuation temperature
is chosen to be very close to the zero-temperature limit, with 
$\gamma T_\chi = 2\times 10^{-9}$ \cite{glatz_2011}.
Beyond these parameters, the
rescaling 
to this normalized form
follows from Ref.~\cite{aranson_2002}.

A TDGL approach
\cite{sademelo_1993,
maly_1999,maly_1997}
can be viewed as a dissipative GP simulation
where the
dissipation
becomes progressively stronger as the dynamics changes from
propagating (in BEC)
to diffusive (in BCS),
reflecting the fact that
pairs are in equilibrium with the
underlying fermions.
Along with this dissipation is a stochastic contribution
required to simulate finite temperatures or experimental imperfections
\cite{glatz_2011}.
Higher levels of microscopic self consistency lead to
longer-lived pairs in the 
intermediate (near-unitary) regime
\cite{maly_1999,maly_1997}, as compared to earlier
estimates of the TDGL coefficients.
This effectively reduces the dissipation (increases $\theta$).
We
use $\theta = 88^\circ$ for simulations here, with the amount of dissipation
primarily affecting the lifetime of the resulting vortex line described below.
[See Supplementary materials (SI) for more details.]

Our studies are based on numerical simulations discretized in
$512^3$ grid points designed to solve Equation
\eqref{eq:tdgl}
using a quasi-spectral split-step method.
The simulations are performed on a GPU computing cluster, allowing relatively
large simulation sizes. The initial condition is
found using a heat diffusion equation
to cool the system to equilibrium before the phase imprinting is applied, as in
Ref.~\cite{glatz_2011}.
Within this TDGL approach for Fermi gases, the form and rescaling of the
trap potential is inserted 
via the local density approximation, 
$\mu \ria \mu - V(\bs x)$, where $\mu$ is the chemical potential [normalized to
1 in Eq.~\eqref{eq:tdgl}]. 
In the bosonic TDGL limit, this trap 
potential form is identical to that used in GP \cite{pethick_2008},
and away from this limit, it is slightly rescaled. 

\textit{Our detailed findings: Transience of defects.}
We investigated a system with axial ($xy$) 
symmetry, $\omega_z = 0.038\mu/\hbar$,
and a variable trap ratio $\lambda = \omega_\perp/\omega_z$. The parameters
used here match the trap ratios and symmetries in Ref.~\cite{yefsah_2013}, but
with a somewhat larger $\omega/\mu$ which provides more fine-grained
computational resolution of the defects.
We will focus on the case 
$\lambda = 3.3$.
In more anisotropic traps with 
$\lambda = 6.2$, we find
the vortex ring is even more transient, immediately colliding with the cloud
edge to leave a line vortex.
Finally, in
quasi-1D clouds  with $\lambda = 15.0$,
the system
cannot nucleate vortices and a reasonably stable soliton
appears which oscillates
along the axial direction of the trap,
in agreement with 1D predictions.

After the cloud has reached equilibrium, 
we imprint a $175^\circ$ phase shift in
the middle of the cloud [in a plane perpendicular to $z$, with axes shown in
Fig.~\ref{fig:r_decay}(a)], and observe the
evolution.
Images from a typical run are displayed in Figure
\ref{fig:r_decay}(b) (see also SI).
Immediately after the imprint, a planar density depletion forms [first frame in
Fig.~\ref{fig:r_decay}(b)]. However, the plane rapidly deforms along the axial
direction (see SI), and a vortex ring forms.
This initial formation of the vortex ring through the snake instability 
\cite{jones_1986,mamaev_1996,
mamaev_1996b,tikhonenko_1996,feder_2000,reichl_2013}
appears in the second frame
of  Figure \ref{fig:r_decay}(b) \footnote{Sound waves are also produced by phase
imprinting, but
dissipate as they reach the trap edge, a key benefit of the dissipation included
in these simulations. This agrees with experimental
observations \cite{yefsah_2013} and is in contrast to the behavior of sound
modes in
GP simulations \cite{bulgac_2014,reichl_2013}.}.  
In the absence of noise and dissipation, 
this ring is
stable and oscillates in the trap, as observed in other simulations
\cite{bulgac_2014,reichl_2013}. 
The presence of
noise perturbs part of the ring, moving it closer
to the boundary.
The
third and fourth frames illustrate the ring collision with the boundary.

\begin{figure*}
\includegraphics[scale=0.40]{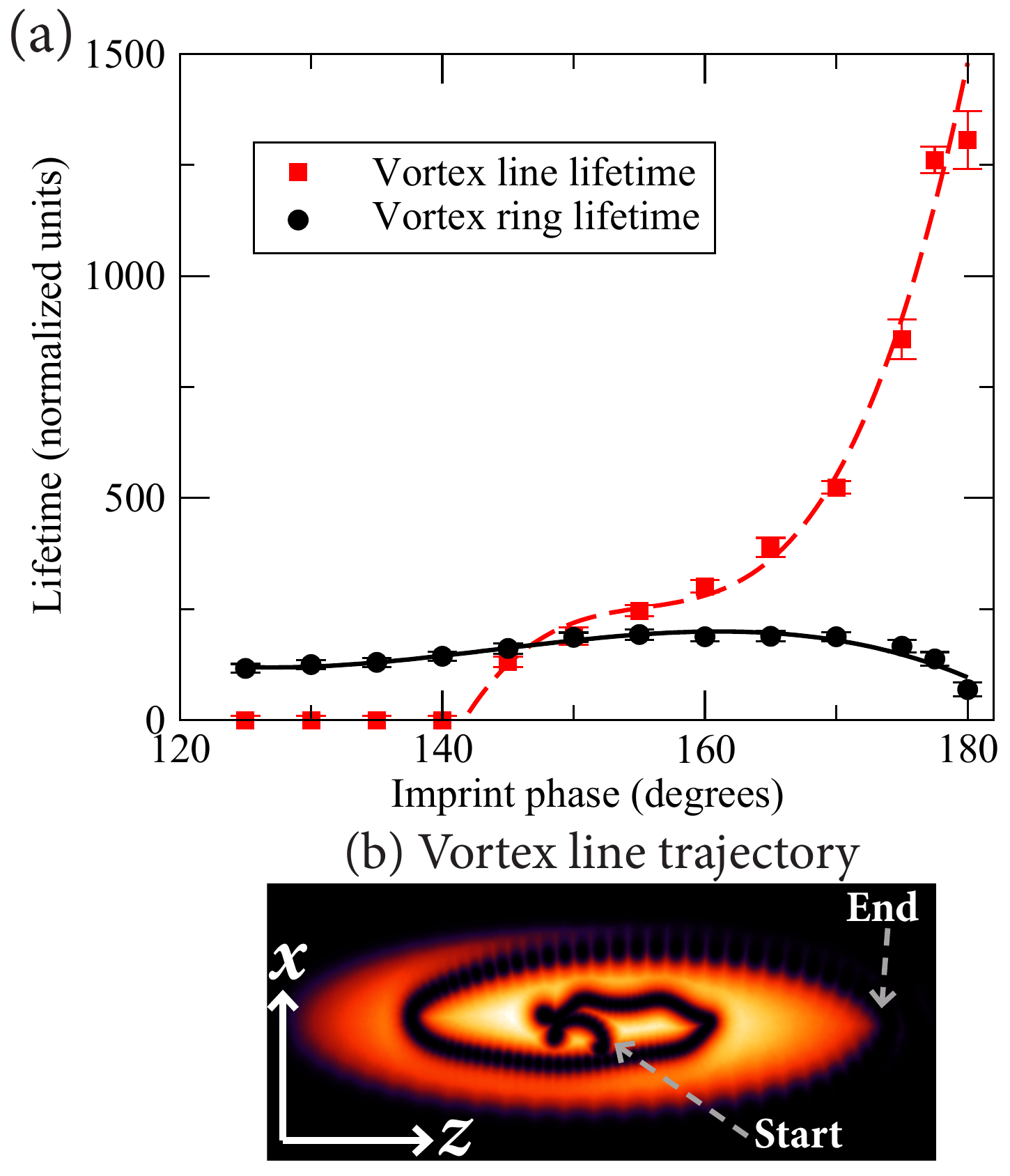}\hspace{0.3in}
\includegraphics[scale = 0.59]
{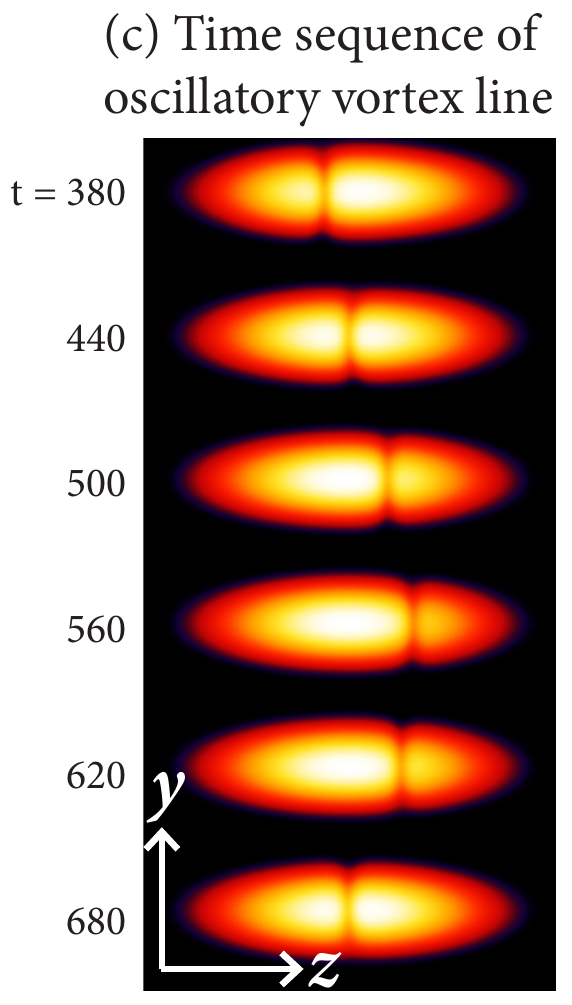}\hspace{0.3in}
\includegraphics[scale = 0.59]{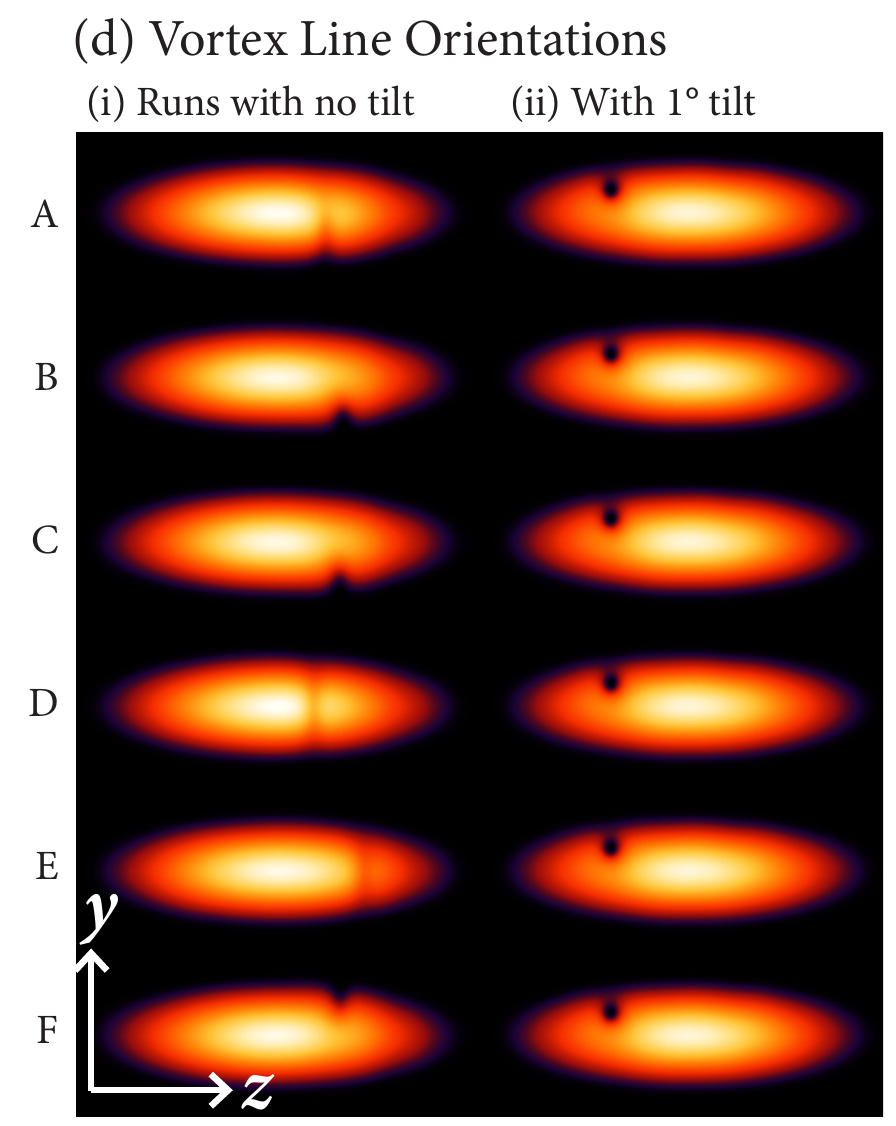}%
\caption{\label{fig:lines_etc}
    (Color online)
(a) Lifetimes of defects as a function of
the imprinting phase,
using the same parameters as in
Figure \ref{fig:r_decay}(b).
Red squares
indicate lifetimes of vortex lines, which are
substantially longer than those of vortex
rings (black circles).
Error bars reflect run-to-run variations
and limitations on time estimates. Lines are cubic polynomial fits provided as a
guide to the eye.
    (b) Plots of the trajectory of the precessing
    single vortex near the end-stage of equilibration for the same run.
    Here frames (cuts of $\abs{\psi}^2$ along the $y = 0$ plane)
    with $\Delta t=5$ were
    combined and the minimum density among all frames selected, displaying the
    vortex trajectory during $t = [180,1160]$.
(c) A series of density plots for the $\lambda = 3.3$ run in Figure
    \ref{fig:r_decay}(b), integrated along the $x$ axis, during the
    interval in which only one vortex is present.  The density effectively shows
    a line depletion due to the orientation of the vortex, which oscillates
    along the cloud.  
    (d) Comparison of a
    (d.i) perfectly orthogonal phase imprinting and (d.ii) phase imprinting with
    a beam tilted by $1^\circ$ relative to the vertical axis, leaving it
    orthogonal
    to the $x$ (short horizontal) axis but not to the $z$ (loosely confined
    horizontal) axis. Each cell is integrated along $x$ at $t = 400$ for a run
    with different random noise, but conditions otherwise
    unchanged. This shows that without a tilt the vortex can be oriented many
    different directions (d.i), but a small tilt stabilizes the
    orientation in different runs (d.ii).
}
\end{figure*}

The observed tilting and attraction of the vortex ring to the boundary
can be qualitatively understood by applying the method of images \cite{schwarz_1985} to
vortex dynamics. (See SI for more details.)
By considering the simplified case of the ring's interaction with a hard planar
boundary, an image vortex ring can be introduced to satisfy the boundary
conditions \cite{schwarz_1985}.
This approximation still captures the main dynamics of the trapped vortex ring 
\cite{mason_2008,mason_2006}.
In this approximation the ring propagates along the boundary due to advection by
supercurrents, but the image vortex retards that
part nearest the boundary leading to a tilt \cite{anglin_2002,
parker_2008}. Finally, the tilted ring is attracted to the boundary, and
``reconnects'' with the image ring, resulting
in a vortex line attached to the boundary [see Figure \ref{fig:r_decay}(b)].
The reconnection 
dynamics are essentially the same as that of the reconnection between two vortex
lines \cite{siggia_1985,schwarz_1985,koplik_1993}.

The stability of these line and ring defects,
and their robustness to unavoidable noise and broken symmetries, are critical to
properly compare simulations to experiment.
In Figure
\ref{fig:lines_etc}(a) we plot the defect lifetime as a function of the
initial phase imprinting angle applied. The vortex ring always decays relatively
rapidly, through shrinking to zero radius at small imprinting angles or from
impacting the trap edge at larger angles. However, at the large phase imprinting
angles appropriate to experiment, the vortex line that remains after ring decay can persist for a
relatively long time, as in Figure \ref{fig:lines_etc}(b-c), where multiple
cycles of the vortex line precession are observed.
These lifetimes are controlled by the advection of the defects, which in turn
depends on 
both the local 
phase gradient
and the superfluid 
density depletion. Similar to solitonic
behavior \cite{busch_2000}, small phase shifts lead to large defect velocities,
while 
phase shifts close to $\pi$ cause large density
depletions and apply very little momentum to the defects. These low velocities
significantly contribute to the long vortex lifetime in this regime
\footnote{This phase-dependence on vortex line velocity reflects the behavior of
``solitonic vortices'' described in recent experiments
\cite{ku_2014,donadello_2014}. In all of these cases, the ``solitonic vortex''
is simply the vortex line with dynamics influenced by
the inhomogeneous, elongated trap it is placed in. Note that this is in contrast
to early discussions of solitonic vortices
\cite{brand_2001,brand_2002,parker_2004}, which focus on quasi-1D systems
which are dissimilar to these recent experiments and simulations.}.

A more detailed view of this
precessing trajectory is shown in Figure \ref{fig:lines_etc}(b).
At the earliest times, when the vortex is near the cloud center, it has a
smaller-scale rotation in the counterclockwise direction in addition to the main
precession of the vortex in the clockwise direction. This smaller-scale
rotation, seemingly due to the decay of the other portion of the ring, gradually
subsides, and the remaining precession and outward movement of the vortex are
typical in the presence of dissipation (see also SI) 
\cite{parker_2008,jackson_1999}.
The precession of this line vortex could appear as a
traveling density depletion. This is illustrated
in Figure \ref{fig:lines_etc}(c), under the assumption that
the vortex line was precessing in a
plane that is nearly orthogonal to the one observed.

\textit{Comparison to Experiment.} 
The results and interpretations of recent experiments \cite{ku_2014} agree well
with the vortex line observations described above.
What remains to be understood, then, is what determines the line vortex orientation. 
As demonstrated by the results of multiple runs in Figure
\ref{fig:lines_etc}(d.i), without further symmetry breaking,
our simulations produce vortex lines with random
orientations. (Similarly, note that the run used in 
Figs.~\ref{fig:r_decay}(b) and
\ref{fig:lines_etc}(b-c) has a random vortex orientation which does not match
experiment.)
In order for the lines to have robust alignment,
as has been found in the most recent experimental analysis \cite{ku_2014}, other
symmetry-breaking features must be present.
One proposed possibility is that asymmetries in the perpendicular trap
frequencies $\omega_x$ and $\omega_y$ govern the vortex orientation
\cite{ku_2014}. 
Another alternative is a slightly tilted
phase-imprinting beam.
With the inclusion of noise in our simulations, we have found that small trap
anisotropies, as are reported experimentally \cite{ku_2014}, 
do not produce reliable vortex orientations in contrast to recent claims
\cite{wlazlowski_2014}. (It should be noted that in our earlier simulations we were inclined
to believe in this mechanism as well, but have since rejected it.) 
The lack of consistency we find reflects the
possibility of reconnection between the 
two vortices that form, and the relatively
chaotic vortex behavior that follows (see SI for details).
This is in contrast to the simulations
\cite{wlazlowski_2014,reichl_2013}
which maintain a 
reflection symmetry about the $yz$ plane ($x \ria -x$), thus leading to
vortex alignments which do not otherwise seem to be stable 
\footnote{
Other recent simulations in BECs
\cite{becker_2013} showed the expected vortex orientation due to anisotropy, 
but only with very
large anisotropy $\omega_y/\omega_x \sim 1.6$.}.

A more satisfactory alternative is to include the effects of very small tilts
in the
phase-imprinting beam, so that it is offset from the vertical axis by an angle
$\alpha$ [see Fig.~\ref{fig:r_decay}(a.ii)] and
no longer perpendicular to the long $z$ axis, while remaining perpendicular to
the $x$ axis. We find this  
can reliably determine the vortex orientation, and agrees with the orientation
observed in experiment. 
As this
imaging beam cannot be placed perfectly, we believe this is a plausible
mechanism for
determining the orientation of the vortex line.
This is shown
in Figure \ref{fig:lines_etc}(d.ii). 
As noted in Ref.~\cite{ku_2014}, slight tilts
of the initial planar depletion can lead to different effects. Here, we find
that for tilts in the imprinting beam of $\alpha = 1^\circ$ 
or smaller, a vortex ring is
produced which quickly decays to a vortex line, as in the absence of
this symmetry breaking. 
The degree of tilt needed to orient the line
is dependent on the magnitude of the noise introduced. For
typical values of our small noise prefactor, an $\alpha \approx 
0.05^\circ$ tilt is adequate.
Furthermore, for a $1^\circ$ tilt we find that moderate anisotropies have no
discernible influence upon the ultimate orientation of the vortex line, though
it may add an additional decay step (see SI).

Thus far, our simulations (with these small broken symmetries) lead us to
believe that what was observed experimentally
\cite{yefsah_2013}
was not a long-lived soliton but rather this residual precessing vortex filament
which can be imaged as in
Figure \ref{fig:lines_etc}(c).
However, in
order to fully compare with experiments, one needs to address the
period of oscillations and the associated trends from the BCS to
BEC sides of unitarity. In general, we expect
that for constant trap parameters, the period of the precession of a vortex
line will follow the size of the superfluid cloud. This means that as the
system approaches unitarity from the BEC side, the increase in interaction
strength will expand the cloud and lead to a longer precession period.
Similarly, if $\omega_\perp$ is kept constant and $\lambda$ decreased, the
weaker $z$-axis confinement will increase the cloud size and lead to a longer
precession period. Both of these trends match the BCS-BEC variation observed experimentally
\cite{yefsah_2013}. 
These predictions can be quantified using the 
Thomas-Fermi (TF) approximation. In this case the 
precession frequency has been derived as \cite{fetter_2001b,
svidzinsky_2000,fetter_2001}
\meq{\label{eq:4}\omega_p =
    \frac{3\hbar\omega_z\omega_\perp}{M(\omega_z^2+\omega_\perp^2)R_{xz}^2}
\log\pth{\frac{R_{xz}}{\xi}}\frac{1}{1-r_0^2},}
where $\xi$ is the healing length, $R_{xz}^2 = 4\mu /
[M(\omega_z^2+\omega_\perp^2)]$
defines an effective trap radius, and $r_0 = (x/R_x,z/R_z)$ is the dimensionless
vortex position. Using the TF approximation
\cite{pethick_2008} for the
parameters $\mu$ and $\omega_\perp$ 
given in Ref.~\cite{yefsah_2013} at unitarity, we obtain a very rough estimate 
of the ratio of the vortex period to the
trap period as $T_v/T_z = 10.2/(1-r_0^2)$, 
in reasonable agreement with the reported results
near unitarity. Our simulations also agree with Equation \eqref{eq:4} (see
SI).

\textit{Conclusion.}
These simulations strongly support the conclusion that earlier experimental
observations \cite{yefsah_2013}, interpreted as stable solitons,
were in fact precessing vortex lines. 
We note that this simulation-based attribution 
of the experimental observations to vortex
lines preceded more recent experiments \cite{ku_2014}. 
Adding to the support for our physical picture is
the fact that 
the oscillation period associated with
the precessing vortex filament
[Eq.~(\ref{eq:4})]
depends directly on the cloud size squared, and that
this period will be longest in the BCS and shortest in the BEC regimes.
Finally, important is the rather robust vortex line alignment we find
with a very small tilt of the phase imprinting angle.

This work demonstrates that in future simulations of superfluid dynamics, the
stability of the results should be demonstrated when imperfections (whether
thermal fluctuations, small anisotropies, misalignments, etc.) are included.
These processes can be necessary for simulations to mimic real-world
experiments, and to properly model the dynamics which lead to ultimate
equilibration. 

%

This work is supported by NSF-MRSEC Grant
0820054.  Work at Argonne was supported by 
the Scientific Discovery through Advanced Computing (SciDAC) program funded
by U.S.\ Department of Energy, Office of Science, Advanced Scientific Computing
Research (large-scale GPU simulations) 
and Basic Energy Sciences, and by the Office of Science, Materials
Sciences and Engineering Division (modeling/analysis). 
The numerical work was performed on NIU's
GPU cluster GAEA. P.S.~acknowledges
support from the Hertz Foundation.  
Finally, we are grateful to William Irvine, Ariel Sommer and Michael Forbes 
for insightful discussions.

\bibliography{Review}
\newpage



\section*{Supplementary material}

## SUPPLEMENTARY MATERIAL

### Detailed Decay of the Defect

A 3D visualization of the main simulation run analyzed in the text is shown in Fig. S1. Sound waves are produced and dissipate at $t = 8$ through $t = 80$. At the same time, the inner core of the planar density depletion "bows out" at $t = 24$, leading to the depletion transforming into a vortex ring ($t = 48$) as fluid returns to the middle of the superfluid. Past this point, the decay of the vortex ring, for example $t = 144$, and the precession of the resulting vortex line from $t = 176$ to $t = 700$ can be observed.

### Method of Images

To qualitatively describe the decay of the vortex ring presented in the main paper, we use the method of images by considering the interactions of a hard boundary which is approximated as a plane. This approach can give an intuitive picture for the movement of the vortex ring and vortex line, and it also draws on techniques shared with electromagnetism. Although this approach is qualitative [1–3] it can provide valuable insight into vortex behavior near boundaries [4]. Importantly, these image ring arguments match our observations in the above simulations.

In more detail, to use the method of images with a hard boundary, one begins with the condition that the normal component of the superfluid velocity is zero on the boundary, $\bm{v}_s \cdot \bm{n} = 0$. In direct analogy with the magnetic field generated by current flowing through a wire, the velocity fields created by vortices in the absence of boundaries follows the Biot-Savart law [5]:

$$\bm{v}_{s,w}(\bm{r}) \equiv \frac{\kappa}{4\pi} \int_{\mathcal{L}} \frac{(\bm{s} - \bm{r}) \times d\bm{s}}{|\bm{s} - \bm{r}|^3} + \bm{v}_{s,i}, \tag{1}$$

where the integral goes over the vortex line(s), $\kappa$ is the quantized vortex circulation, and $\bm{v}_{s,i}$ is due to imprinting. However, to satisfy the boundary condition $\bm{v}_s \cdot \bm{n} = 0$, a second (boundary) velocity field $\bm{v}_{s,b}$ is added, with $\nabla \cdot \bm{v}_{s,b} = 0$ and $\nabla \times \bm{v}_{s,b} = 0$, such that

$$(\bm{v}_{s,w} + \bm{v}_{s,b}) \cdot \bm{n} = 0. \tag{2}$$

For a plane boundary, $\bm{v}_{s,b}$ is the field created by an image vortex, which is the real vortex reflected along the boundary with the vortex direction reversed [5].

A diagram of this image vortex behavior is presented in Figure S2. As can be seen in this figure, and as described in the main text and seen in simulations, the boundary has two major effects. First, it retards the movement of the portion of the vortex ring nearest the boundary, resulting in a tilt of the ring as a whole. Second, because vortices of opposite direction attract each other, the ring is attracted to its image, and eventually reconnects with it. Thus, the use of images, controlled by application of boundary conditions and the Biot-Savart law, leads directly to a qualitative description of the ring decay observed in these simulations.

### Vortex Decay in Anisotropic Traps

Recent experimental work [6] has postulated that the reliable vortex orientations produced are due to a slight difference in the trapping frequencies $\omega_x$ and $\omega_y$, primarily due to the force of gravity decreasing the effective $\omega_y$. In this trap configuration, the vortex is in its lowest energy state if oriented parallel to the $x$ axis, which would agree with experiment. However, recent work [7] has pointed out that the relaxation of the vortex line to this orientation may be quite slow. While Ref. [7] shows other potential mechanisms that result in this vortex decay, as we note in the main text, their simulations depend on maintaining a reflection symmetry.

In contrast to these suggestions, we find that the vortex orientation in an anisotropic trap can be very sensitive both to the details of the experimental configuration, as well as to random perturbations of the system. Some of these features are explored in Figure S3(A-C). More specifically, Fig. S3(A) shows a large system in which the ring impacts both sides of the trap, and breaks apart into two vortices. However, after a period of time these two vortices reconnect, a fairly chaotic process that rapidly ejects one vortex line, leaving a single vortex line in the trap. This remaining vortex line can have a fair degree of freedom in its orientation, as shown in $t = 850$ for (A). In Fig. S3(B), the ring similarly splits into two vortices, but these vortices move far enough apart that they do not reconnect, similar

to the case for large anisotropies reported in Ref. [9]. This produces the desired vortex orientation, but we do not observe it occurring reliably. Fig. S3(C) is similar to (A), but for a smaller system, and is a very common decay mechanism that we observe in anisotropic traps, with a relatively free final vortex orientation.

When a tilt is applied in the presence of an anisotropic trap, we find that the tilt appears to determine the vortex orientation. In Fig. S3(D) a 1° tilt is applied perpendicular to the $x$ axis. The vortex ring does break apart into two vortex lines, but these two vortex lines reliably avoid reconnection, with one of the two quickly leaving the trap. We note that this tilt of the beam seems reasonable given the fact that the phase imprinting beam is sent from above the cloud [10], so it could be slightly tilted away from parallel to the $y$-axis, while still applying a nearly symmetric cut of the cloud along the $x$ direction.

In contrast, in Fig. S3(E) the tilt is applied perpendicular to the $y$ axis (opposite what might be seen in experiment), and the final vortex orientation follows the tilt rather than the trap anisotropy. Specifically, in (E) the initial vortex ring quickly collides with one edge of the trap, leaving a vertically oriented vortex line.

In summary, we find that some mechanisms of vortex ring decay in an anisotropic trap do reproduce the vortex orientation seen experimentally [6]. However, when noise is included, we do not find that these orientations are robust. Instead, if a tilt in the imprinting beam is present, the tilt robustly dictates the vortex line orientation.

### Periods and Dissipation

The periods which we measure for our simulations seem to agree well with theoretical predictions. For our simulations (with smaller $\mu/\omega_\perp$ values compared to the experiments), Eq. (2) of the manuscript predicts $T_v/T_z = \{3.5, 2.3\}(1 - r_0^2)$ for $\lambda = \{3.3, 6.2\}$ respectively (with $T_v$ the vortex line period, and $T_z = 2\pi/\omega_z$). We measured an average $T_v/T_z = \{2.4 \pm 0.3, 1.6 \pm 0.2\}$ for these $\lambda$, which agree for $r_0 \approx 0.7$. We also observed the predicted decrease in period as the vortex moves to larger radii. Furthermore, although the vortex ring exhibited the slowest oscillation of the defects considered (we reduced noise to 0 and observed $T_{vr}/T_z = 1.8 \pm 0.2$ for $\lambda = 6.2$), the distinction between it and the line vortex is not large, and we do not consider the oscillatory period found in the experiment to point to definitive proof of a vortex ring [11]. Finally, for the quasi-1D case with $\lambda = 15$, we measure the soliton period $T_s$ as $T_s/T_z = 1.39 \pm 0.03$, in good agreement with the predicted $T_s/T_z = \sqrt{2}$ value for solitons in the GP system.

Figure S4 shows the effect of changing the dissipation angle $\theta$ in these simulations. At small angles, the high dissipation causes the vortex line to very quickly leave the superfluid. In contrast, at large angles, where dissipation is very small, the vortex can precess for many cycles around the cloud, and moves toward the edge very slowly. This is in agreement with the theoretical expectations for the effect of dissipation [12].

Also available are movies [13] which show integrated cloud densities at $\theta = 86°$, $88°$, and $89.5°$. In the movie with $89.5°$, two effects of low dissipation can be seen: Non-topological density fluctuations (i.e. sound waves) persist for a long period of time, and the vortex precesses for many cycles before dissipating at the trap edge.

[1] P. Mason and N. G. Berloff, Phys. Rev. A **77**, 032107 (2008).
[2] P. Mason, N. G. Berloff, and A. L. Fetter, Phys. Rev. A **74**, 043611 (2006).
[3] J. R. Anglin, Phys. Rev. A **65**, 063611 (2002).
[4] While Ref. [3] argues against the use of image vortices for quantitative work in trapped gases, the author does acknowledge their qualitative predictions are usually accurate. Furthermore, Ref. [1] is able to use image vortices to analyze essentially the same problem.
[5] K. W. Schwarz, Phys. Rev. B **31**, 5782 (1985).
[6] M. J. H. Ku, W. Ji, B. Mukherjee, E. Guardado-Sanchez, L. W. Cheuk, T. Yefsah, and M. W. Zwierlein, arXiv:1402.7052 [cond-mat.quant-gas] (2014).
[7] G. Wlazłowski, A. Bulgac, M. M. Forbes, and K. J. Roche, arXiv:1404.1038 [cond-mat.quant-gas] (2014).
[8] T. Yefsah, A. T. Sommer, M. J. H. Ku, L. W. Cheuk, W. Ji, W. S. Bakr, and M. W. Zwierlein, Nature **499**, 426 (2013).
[9] C. Becker, K. Sengstock, P. Schmelcher, P. G. Kevrekidis, and R. Carretero-Gonzlez, New J. Phys. **15**, 113028 (2013).
[10] A. Sommer, Private communication (2014).
[11] Note that for this run, double precision was used but with a smaller computational grid of edge size $L = 256$ rather than $L = 512$ used in the rest of the paper.
[12] N. G. Parker, B. Jackson, A. M. Martin, and C. S. Adams, in *Emergent Nonlinear Phenomena in Bose-Einstein Condensates*, Atomic, Optical, and Plasma Physics No. 45, edited by P. P. G. Kevrekidis, P. D. J. Frantzeskakis, and P. R. Carretero-González (Springer Berlin Heidelberg, 2008) pp. 173–189.
[13] Available at http://mti.msd.anl.gov/highlights/coldfermions.

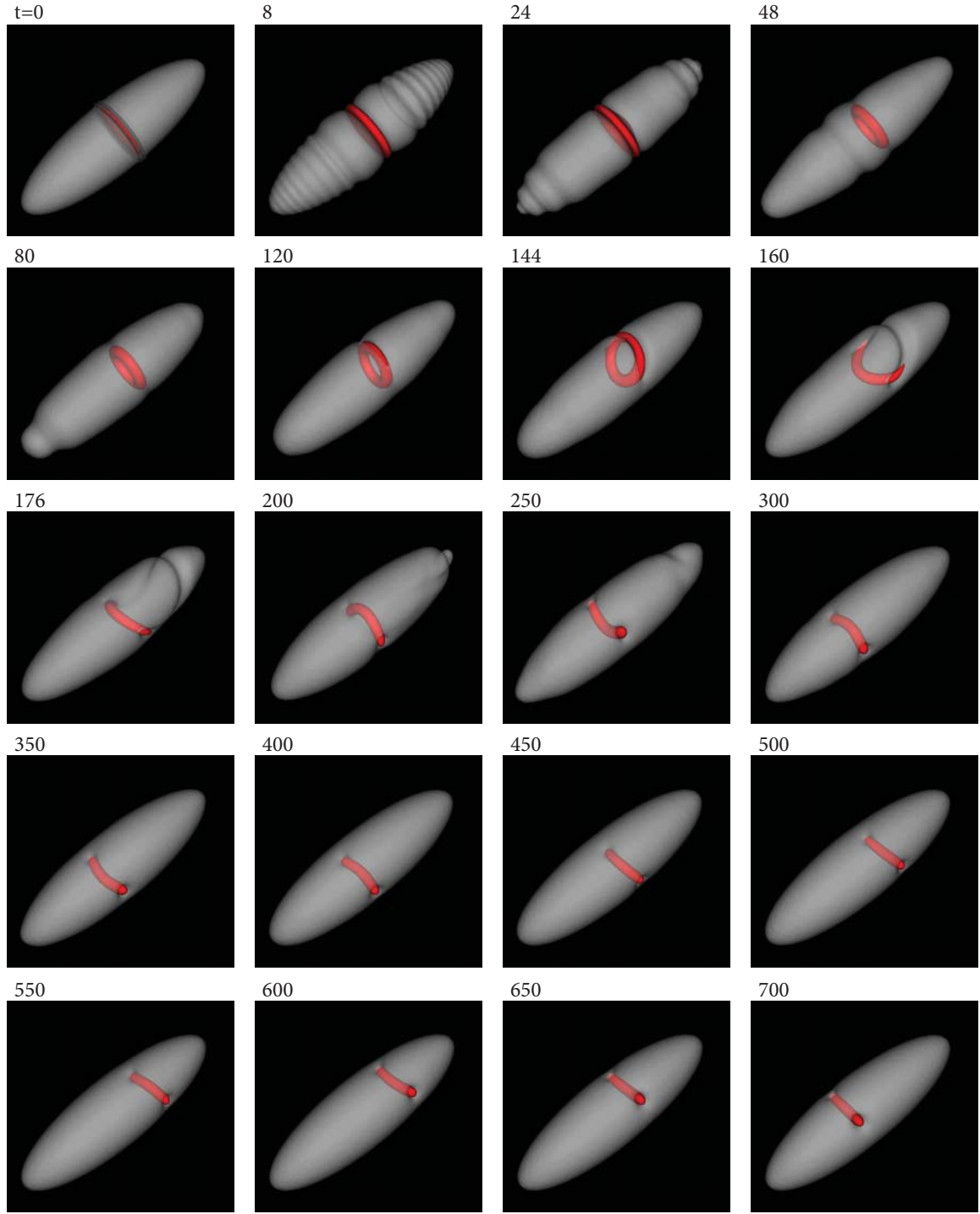

FIG. 1. A time sequence of 3D images for the run in Figure 1(b) at the indicated times. Each timestep includes an isosurface of $|\psi|^2$ (white) and of $|\psi_e|^2 - |\psi|^2$ (red). Here $|\psi_e|^2$ is the superfluid amplitude measured just prior to the phase cut, when the cloud is in equilibrium.

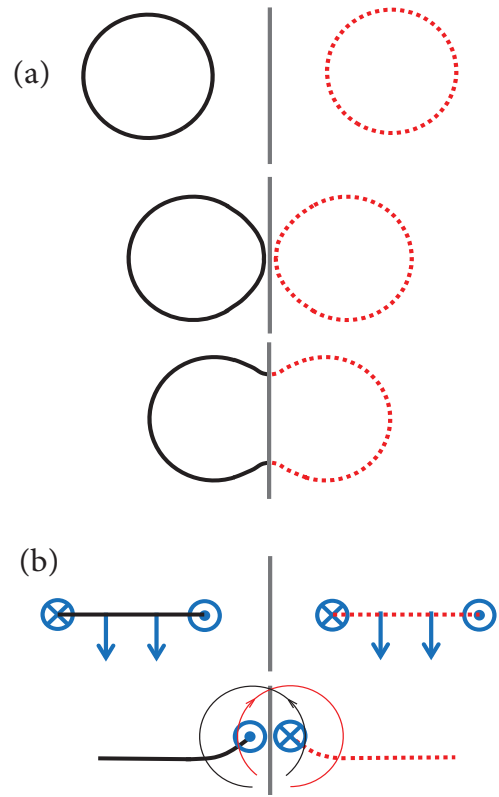

FIG. 2. Illustration of a vortex ring (black) to image ring (dotted red) reconnection process [5], from top to bottom and viewed along the $z$ axis in (a). Panel (b) shows the side view parallel to the boundary plane and vortex ring plane. Circulation corresponds to $\{\odot, \otimes\}$ for counterclockwise and clockwise rotation, respectively, in this view. Arrows indicate velocities induced by the vortex ring segments nearest the surface.

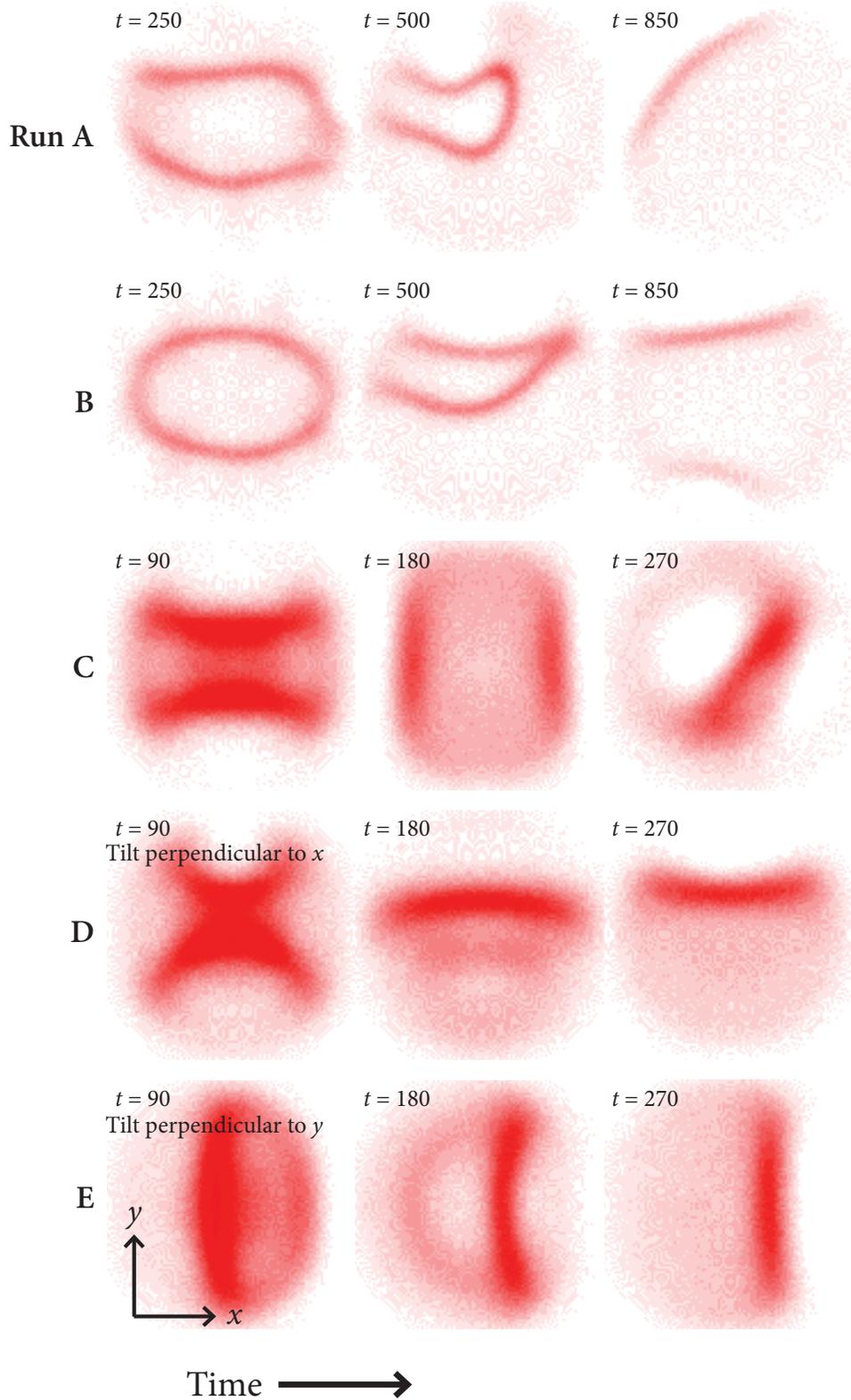

FIG. 3. Results of simulations runs in anisotropic traps. Each row shows a time sequence of vortex evolution in one simulation run. Displayed are residuals of the cloud density, integrated along the $z$ (long horizontal) axis. The red color represents regions of the integrated cloud which are less dense than the density for the equilibrated cloud. All clouds here have $\omega_y < \omega_x$, with $\omega_y/\omega_x \approx 0.96$, which reflects both the direction and magnitude of anisotropy reported in recent experiments [6]. In these figures, the experimentally reported vortex lines [6, 8] would appear as horizontal red lines. Here (A-C) are runs with no tilt of the imaging beam applied, while (D) and (E) do have a tilt applied, as given in the figure. In (A) and (B) all trap frequencies have been reduced by a factor of two to give a larger system size, and the effective temperature for the noise, $T_\chi$, has been decreased by a factor of about 12 compared to all other runs in this manuscript.

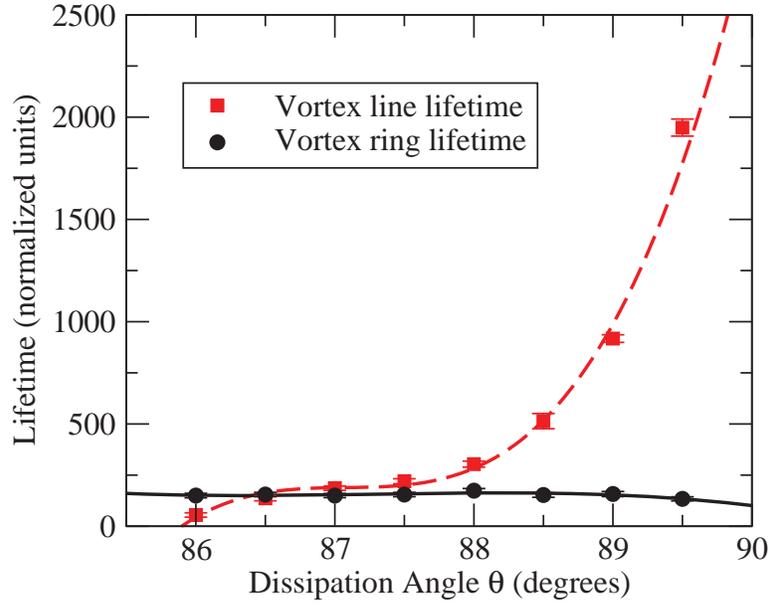

FIG. 4. The lifetime of defects as a function of the dissipation angle $\theta$ used in the simulations. The vortex ring lifetime (black circles) is relatively consistent with the angle, while large values of $\theta$ give very large vortex line lifetimes (red squares). Lines are cubic polynomial fits provided as a guide to the eye. Note that rather than a step function phase imprinting, these simulations used a phase imprinting formula of $\Delta\phi = \frac{1}{2}\left[1 + \tanh(-x/\delta)\right]$ with $\delta = 0.1$ in normalized units. This causes lifetimes here to differ somewhat compared to Fig. 2(a) of the manuscript



\end{document}